\documentclass[reprint,prl,footinbib,twocolumn,superscriptaddress,lettersize,floatfix]{revtex4-1}

\usepackage{graphicx}
\usepackage{bm}
\usepackage{hyperref}
\usepackage{bbm}
\usepackage[letterpaper, portrait, margin=1in]{geometry}
\usepackage{color}
\usepackage{comment}

\begin{document}

\title{Deterministic generation of large-scale entangled photonic cluster state from interacting solid state emitters}

\author{Mercedes Gimeno-Segovia}
\affiliation{Quantum Engineering Technology Labs, H. H. Wills Physics Laboratory and Department of Electrical and Electronic Engineering, University of Bristol, BS8 1FD, UK}
\affiliation{Department of Physics, Imperial College London, London SW7 2AZ, UK}
\author{Terry Rudolph}
\affiliation{Department of Physics, Imperial College London, London SW7 2AZ, UK}
\author{S. E. Economou}\email{economou@vt.edu}
\affiliation{Department of Physics, Virginia Tech, Blacksburg VA 24061, USA}

\begin{abstract}
The ability to create large highly entangled `cluster' states is crucial for measurement-based quantum computing. We show that deterministic multi-photon entanglement can be created from coupled solid state quantum emitters without the need for any two-qubit gates and regardless of whether the emitters are identical. In particular, we present a general method for controlled entanglement creation by making direct use of the always-on exchange interaction, in combination with single-qubit operations. This is used to provide a recipe for the generation of two-dimensional, cluster-state entangled photons that can be carried out with existing experimental capabilities in quantum dots.
\end{abstract}

\maketitle

The cluster state quantum computing paradigm is believed to be the most feasible approach for photonic quantum computing. In this approach, the difficulty of entanglement creation between photons is shifted to the upfront creation of a highly entangled multiqubit cluster state \cite{Raussendorf_PRL01}. To date, photonic cluster states have been created by passing parametrically down-converted pairs of entangled photons through linear optic elements and subsequent measurement of a photon. This process is inherently probabilistic, and as a result creating a cluster state larger than a few photons is a formidable task \cite{Wang_PRL16b}. In previous theoretical work it was shown \cite{Lindner_PRL09} that a periodically pumped quantum dot (QD) can produce a cluster state string of photons. Very recently, there has been an experimental breakthrough materializing this deterministic approach and generating a one-dimensional cluster state \cite{Schwartz_Science16}. However for applications, larger dimensional graph states are needed. To that end, a proposal \cite{Economou_PRL10} generalized the scheme of Ref. \cite{Lindner_PRL09} to a pair of QDs, introducing the idea that entangled emitters can emit entangled photons. The main challenges with that approach are that it requires the application of experimentally demanding two-qubit entangling gates between the emitters, and that it assumes that the two QDs do not interact in the absence of optical pulses. Although there are ongoing efforts to demonstrate this idea, these issues make the experiments challenging. The recent experimental progress of Ref. \cite{Schwartz_Science16} makes a practical protocol to generate a higher dimensional cluster state with existing resources a particularly timely topic. In addition to quantum computing, the deterministic creation of large-scale cluster states would also impact quantum communications \cite{Buterakos_PRX17,Russo_prb18}.

Here we present a deterministic protocol for generating two-dimensional photonic cluster states which requires no externally driven two-qubit gates and which allows for, and in fact makes crucial use of, an always-on coupling between emitters. The necessary entanglement is built up by free evolution and the photons are generated in an entangled state through optical pumping of the emitters. For QDs it is the always-on exchange interaction between the spins that provides entanglement. Remarkably, we show that with carefully chosen pulse sequences this entanglement is sufficient to generate a cluster state when combined with single-qubit gates already demonstrated in experiments. We provide the pulse sequences that implement the required evolution for two distinct cases, emitters with (i) equal and (ii) unequal Zeeman splittings.

While the scheme we describe is applicable to any pair of emitters coupled with Heisenberg type interaction, we focus on QDs, because they are very efficient emitters, and the coupled-QD system has been studied and understood very well experimentally \cite{Krenner_PRL05,Stinaff_sci06,Doty_PRB08,Kim_NP11}. We consider a pair of stacked epitaxial QDs with a thin enough barrier in between such that they are tunnel coupled. A bias voltage controllably loads single carriers (electrons or holes) into each QD.  We consider the bias regime where each QD contains a single electron. The electrons can virtually tunnel into the opposite QD, and thus there is an effective exchange interaction between them (Fig. 1). Recent experimental advances based on this system demonstrated ultrafast coherent control, including single spin rotations and entanglement control \cite{Kim_NP11}.

In previous work \cite{Economou_PRL10}, we showed that to generate a photonic cluster state the required evolution of the two emitters should have the form $U_{target}=CZ (A\otimes A)$, where $CZ$ is the conditional-Z gate between the two spins, given by the matrix $diag(1,1,1,-1)$ and $A$ is either the Hadamard gate, $H$, or an equivalent gate, e.g., a $\pi/2$ rotation about an axis in the $xy$ plane. Successive applications of $U_{target}$, each followed by optical excitation and spontaneous photon emission results in a two-dimensional ladder of entangled cluster state photons. In Ref. \cite{Economou_PRL10} the suggested $CZ$ gate was implemented optically by exploiting an electron-hole exchange interaction \cite{Economou_PRB08} in the excited state and the two QDs were assumed completely decoupled in their ground states.

The growth axis  ($z$ axis) of the two QDs is a preferred direction along which the symmetry is lowered and which also coincides with the laser propagation direction. Due to the broken symmetry along $z$, there are optical polarization selection rules associated with that axis. As a result, the most straightforward operations to implement are initialization, measurement and spin rotations about $z$ \cite{Press_Nat08,Greilich_NP09}. This is due to the fact that polarization alone provides selectivity between spin states along $\pm z$, eliminating the need for frequency selectivity which would necessitate longer pulses. We consider a magnetic field $B$, which for simplicity we fix perpendicular to the $z$ axis, defining the $x$ axis.
\begin{figure}[htp]
	\centering
	\includegraphics[width=1\columnwidth]{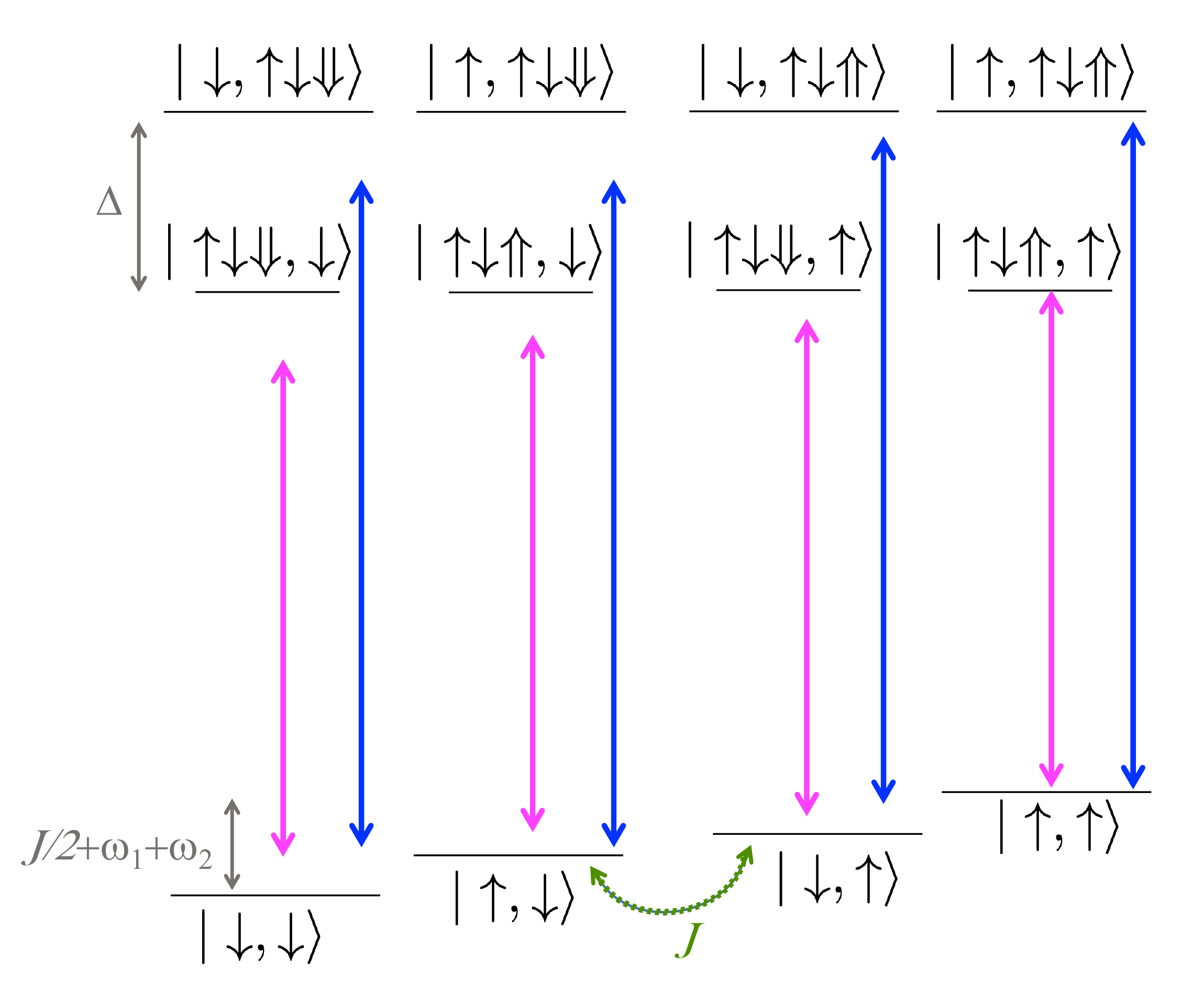}
	\caption{Energy levels of QDs with always-on spin exchange. The difference between the optical transitions in the two QDs is $\Delta\gg J$. The two middle states in the ground state manifold are not energy eigenstates; they are coupled by $J$, as indicated by the arrow.}
	\label{qdlevels}
\end{figure}

In practice, two-spin experiments \cite{Kim_NP11} are conducted in a regime where the tunnel-coupling in the ground state is strong, resulting in an always-on interaction between the spins, while the excited states are significantly detuned, making the inter-dot interaction in the excited state practically zero. The Hamiltonian in the absence of pulses is $H_0{=}J s_1{\cdot} s_2 {+} \omega_1 s_{1x} {+} \omega_2 s_{2x}  $, where $\omega_j=g_j B$ is the Zeeman frequency for dot $j$, $J$ is the exchange interaction strength, and $s_{ij}$ is the spin matrix of qubit $i$ along the axis $j$. In this regime, single-spin operations, in particular measurement and rotations about $z$, are simple to implement and have been demonstrated. The relevant QD levels are shown in Fig.~\ref{qdlevels}. The ground states have one excess electron per QD in the conduction band; the excited states have an electron in one QD and a trion (two electrons in a singlet in the conduction band and a hole in the valence band) in the other. There is generally flexibility with engineering or tuning QD parameters. The exchange interaction $J$ is determined by the overlap of the electronic wavefunctions in the two QDs; it can be modified by changing the barrier height between the QDs \cite{Krenner_PRL05,Stinaff_sci06,Doty_PRB08}. Ref. \cite{Doty_PRB08} in particular studies the properties of doubly charged coupled QD pairs, which are the focus here. 

The objective here is to generate a 2D cluster state using \emph{only} the always-on exchange interaction in combination with single qubit operations, elements that are attainable with existing experimental capabilities \cite{Kim_NP11}. Different QDs will generally have different $g$ factors. There is correlation between the value of the $g$ factor and the size and composition of the QD and thus some control can be achieved on this value during growth. Given this flexibility, and the possibility that the $g$ factors of a stacked QD pair may not be identical, we examine separately the equal and unequal Zeeman splitting case. The symmetry of the problem is different in each case, resulting in evolution operators of distinct symmetries and thus different pulse sequences. In each case we decompose the evolution operator using the Cartan decomposition \cite{Kraus_PRA01} into a product of single-spin operations and purely two-spin operations. This allows us to identify the parameter regime that maximizes the entanglement generated between the two spins and to isolate the purely single-spin part of the evolution, which may be used or may need to be compensated with single-qubit gates.

\noindent\emph{Equal Zeeman frequencies}--In this case the evolution operator can be decomposed into the form
\begin{eqnarray}
U_{eq} =  \left( e^{-i \omega t s_x} \otimes e^{-i \omega t s_x} \right) e^{-i J t \mathbf{s}_1\cdot \mathbf{s}_2 }.
\end{eqnarray}
The purely two-spin operator in this equation comes from the exchange interaction. It is well-known that Heisenberg exchange interaction yields an entangling gate, the so-called square-root of SWAP, $U_{ss}$. After evolution time $t=\pi/(2J)\equiv\tau_1$, $U_{eq}$ is equivalent to $U_{ss}$ up to single-qubit rotations and is thus maximally entangling:
\begin{eqnarray}
U^{me}_{eq} &=&  \left( e^{-i \frac{\pi\omega}{2J} s_x} \otimes e^{-i \frac{\pi\omega}{2J} s_x} \right) e^{-i \frac{\pi}{2} \mathbf{s}_1\cdot \mathbf{s}_2},\nonumber
\\
&\equiv &  \left( e^{-i \frac{\pi\omega}{2J} s_x} \otimes e^{-i \frac{\pi\omega}{2J} s_x} \right) U_{ss}.
\label{umaxenteq}
\end{eqnarray}
This evolution operator can be used to generate the target gate $U_{target}=CZ(A\otimes A)$ in the case where $A$ is a $\pi/2$ rotation about $x$. To see this, first express $CZ$ in terms of $U_{ss}$ (up to a global phase):
\begin{eqnarray}
\text{CZ} = \left( e^{-i \frac{\pi}{2} s_z} \otimes e^{i  \frac{\pi}{2} s_z} \right) U_{ss}\left( \mathbbm{1} \otimes e^{-i \pi s_z}   \right)U_{ss} .
\label{czdecomp}
\end{eqnarray}
The target gate is
\begin{eqnarray}
U_{target} = \text{CZ}( e^{-i \frac{\pi}{2} s_x} \otimes e^{-i \frac{\pi}{2} s_x}). %=\nonumber
%\\
%e^{i \frac{7\pi}{4}} \left( e^{-i \frac{\pi}{2} s_z} \otimes e^{i \frac{\pi}{2} s_z} \right) U_{ss}\left( e^{i \pi s_z} \otimes \mathbbm{1}  \right)U_{ss}( e^{-i \frac{\pi}{2} s_x} \otimes e^{-i  \frac{\pi}{2} s_x})
%\nonumber
\label{utargeteq}
\end{eqnarray}
Using Eqs.~(\ref{umaxenteq})-(\ref{utargeteq}) we have
\begin{eqnarray*}
U_{target} =
\left( e^{-i \frac{\pi}{2} s_z} \otimes e^{i \frac{\pi}{2} s_z} \right) \left( e^{-i \frac{\pi\omega}{2J} s_x} \otimes e^{-i \frac{\pi\omega}{2J} s_x} \right)U^{me}_{eq} \times  &&\nonumber
\\
\times\left( \mathbbm{1} \otimes e^{-i \pi s_z}  \right) \left( e^{-i \frac{\pi}{2}(\frac{\omega}{J}+1) s_x} \otimes e^{-i \frac{\pi}{2}(\frac{\omega}{J}+1) s_x} \right)U^{me}_{eq}. &&
\label{utargetfinal}
\end{eqnarray*}
Thus, for equal Zeeman frequencies, we may generate the target gate using two sets of single-qubit gates interspersed with periods of free evolution. 

\noindent\emph{Unequal Zeeman frequencies}--The unequal Zeeman splitting case is somewhat more involved, as the symmetry of the system is lower. As a result, the evolution operator does not have a simple decomposition as in the equal Zeeman case above. The evolution operator in the product spin basis is symmetric, with $u_{24}= u_{13}$, $u_{33}= u_{22}$, and  $u_{44}= u_{11}$. The expressions for the matrix elements are in \cite{supplement}.

A key difference from the equal Zeeman case is that the condition for maximal entanglement, in addition to $J$, also depends on $\omega_1,\omega_2$. When
\begin{eqnarray}
t = \frac{2n\pi}{\sqrt{J^2 +(\omega_1-\omega_2)^2}} = \frac{(2m+1)\pi}{J} \equiv \tau_2
\label{tcond1}
\end{eqnarray}
with $n$, $m$ positive integers, the free evolution amounts to an Ising gate up to single qubit operations:
\begin{eqnarray}
U^{me}_{uneq} &=&  \left( e^{i  \phi s_x}\otimes  e^{i \phi s_x}\right)  e^{-i \pi s_x\otimes s_x },
\label{Umeuneq}
\end{eqnarray}
with
\begin{eqnarray}
\phi=\pm\frac{\pi}{2}\sqrt{4n^2-(2m+1)^2} \frac{(\omega_1+\omega_2)}{|\omega_1 -\omega_2|}\pm k\pi \label{phi},
\end{eqnarray}
with $k$ integer from which we obtain the constraint
\begin{eqnarray}
n>|1+2m|/2.
\label{nmcondition}
\end{eqnarray}
Condition (\ref{tcond1}) requires choosing the integers $n$ and $m$ such that the values of $\omega_1$, $\omega_2$, and $J$ fall in a physical range.

To construct the sequence that will give us the target evolution we make an ansatz using single spin rotations about the $z$ and the $x$ axes in addition to the maximally entangling free evolution $U^{me}_{uneq}$. In fact we notice that the square of $U^{me}_{uneq}$ gives a separable evolution of the two spins amounting to rotations about the $x$ axis for both spins. This is a resource we also use as it will provide additional $x$ rotations without the need of external pulses. We equate our ansatz sequence
\begin{eqnarray}
U_{ans}&=& e^{i\alpha} \left( e^{i \phi_{z2} s_z}\otimes  e^{i \phi_{z2} s_z}\right)\left(U^{me}_{uneq}\right)^2 \times \nonumber
\\
& &\times\left(  e^{i \phi_{x} s_x}  \otimes   e^{i \phi_{x} s_x}\right)
 \left( e^{i \phi_{z1} s_z}\otimes  e^{i \phi_{z1} s_z}\right) U^{me}_{uneq},
\nonumber
\label{ansatz}
\end{eqnarray}
to the target gate, $\text{CZ}(H{\otimes} H)$, and examine whether this equation has a general solution and in that case determine the angles $\phi_{z1},\phi_{z2},\phi_{x}$. Multiplying both sides on the right by $(H\otimes H)$ and inserting the identity before the entangling part of $U^{me}_{uneq}$ in the form $(H{\otimes }H)^2$ we notice that we can introduce $CZ$ on the LHS, since $(H{\otimes} H)e^{i \pi s_x\otimes s_x }(H{\otimes} H)$ is proportional to $CZ$ up to single-qubit rotations. We thus  reduce the problem to a single spin equation:
\begin{eqnarray}
e^{i s_z \phi_{z2}} e^{i s_x (2\phi-\pi)} e^{i s_x \phi_{x}} e^{i s_z \phi_{z1}} e^{i s_x \phi} H e^{i s_z \frac{-\pi}{2}} = \mathbbm{1}.
\label{singlespineqn}
\end{eqnarray}

In the middle of the expression in Eq.~(\ref{singlespineqn}) the rotations about the $x$ axis are left separate to emphasize that one originates from free evolution while the other one is implemented by external fields. The solution of Eq. (\ref{singlespineqn}) is
\begin{eqnarray}
\phi_{z1} = \frac{\pi}{2}, ~ 
\phi_{z2} = \pi-\phi, ~
\phi_{x} = \frac{3\pi}{2}-2\phi, ~
\alpha = \frac{5\pi}{4}. \nonumber
\end{eqnarray}
We thus find that the overhead for generating each pair of entangled photons is two single-qubit $z$-rotations and a single-qubit $x$-rotation.
\begin{figure}[htp]
	\centering
\includegraphics[width=1\columnwidth]{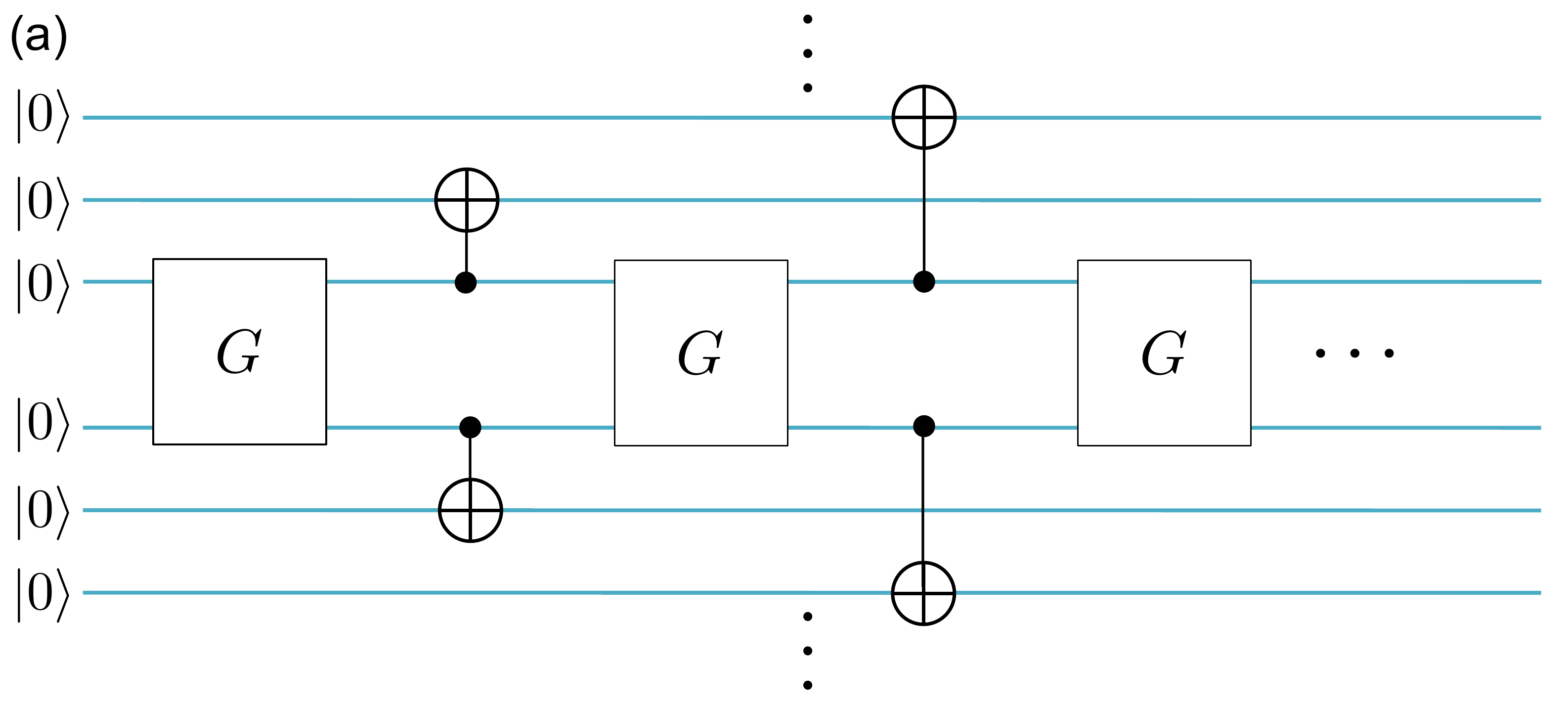}
\includegraphics[width=1\columnwidth]{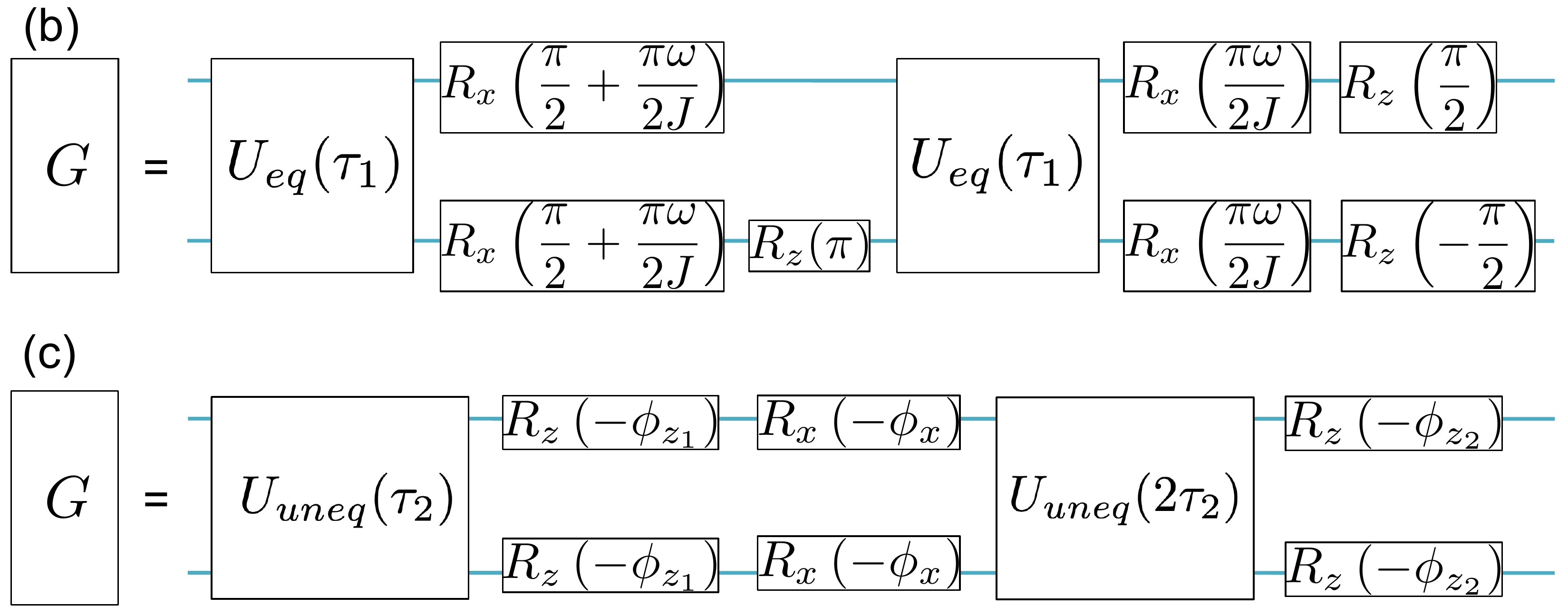}
\caption{(a) Circuit generating entangled photons from QDs, with the spin-spin gate $G$ shown for (b) equal Zeeman frequencies and (c) unequal Zeeman frequencies. }
	\label{circuit}
\end{figure}
The protocols for generating the cluster states for equal and unequal Zeeman frequencies are shown in Fig.~\ref{circuit}.

\emph{Implementing the single-qubit x rotations}--For the $x$ rotations we have several options; using longer pulses, i.e., spectral selectivity; combining free evolution with $z$ rotations; specially engineering or selecting QD parameters to simplify the evolution. Because our primary goal is to find protocols based on experimentally demonstrated capabilities, we will focus on creating the $x$ rotation by combining $z$ rotations and free evolution. Below we discuss this approach in detail both for the equal and the unequal Zeeman frequency cases as applicable.

Free evolution for appropriate time interval $t_x$ (for the equal Zeeman case, $t_x{=}2\pi/J$ and for the unequal $t_x {=}{4n\pi}{/}{\sqrt{J^2 {+}(\omega_1{-}\omega_2)^2}}$) yields an $x$ rotation by angle $\chi$ for each qubit (for the equal Zeeman case, $\chi{=}{-}2\pi\omega/J$ and for the unequal $\chi {=}2\phi$, with $\phi$ in Eq. (\ref{phi})). Combining this with the ansatz
\begin{eqnarray}
%e^{-i\varphi s_x}=e^{-i(\frac{\pi}{2}+\xi) s_z} e^{-i\chi s_x} e^{-i(3\pi+\psi) s_z} e^{-i\chi s_x} e^{-i(\frac{\pi}{2}+\xi) s_z}
e^{i\varphi s_x}=e^{-i\xi s_z} e^{i\chi s_x} e^{-i\psi s_z} e^{i\chi s_x} e^{-i\xi s_z},
\label{xrotfromz}
\end{eqnarray}
which is satisfied for certain ranges of $\chi$, limiting the ratio of the physical parameters to certain values, allows us to tune the $x$-rotation angle $\varphi$ by adjusting the $z$-rotation angles $\xi$ and $\psi$. Out of the physically viable ranges for the system parameters ($\omega$ and $J$), in the equal-Zeeman case we select the large range $\omega  {\in } [0.58J,0.87J]$, a condition that can be achieved by tuning the magnetic field. In the unequal Zeeman case, Eq. \ref{xrotfromz} gives a condition on the ratio of the g factors of the two emitters. The relevant range here is much narrower, so an alternative approach may be more desirable. By selecting parameters appropriately, we can avoid the $x$ rotation altogether in the pulse sequence forming $U_{ans}$. By combining the two middle $x$ rotations in Eq. (\ref{singlespineqn}) we see that in the special case when $\phi{=}3\pi/4$ the externally induced $x$ rotation is not needed ($\phi_x{=}0$). This condition gives us a relation between the Zeeman frequencies of the two QDs through the constraint
\begin{eqnarray}
\omega_1 = \frac{\gamma+1}{\gamma-1}\omega_2,~~\gamma \equiv \frac{\pm 3/2 \pm 2k}{2\sqrt{4n^2-(2m+1)^2}},
\end{eqnarray}
with $k$ a positive integer. A large number of solutions can be obtained by varying $n,m,k$. More importantly, a reasonably large number of solutions persists for physically relevant parameter regimes, $\omega_1/\omega_2{\in}(1, 1.2)$. For example, $n{=}14,m{=}13,k{=}45$ gives $\gamma{=}183/(2\sqrt{55})$ so that $\omega_1/\omega_2{\approx}1.18$, which should be achievable experimentally, e.g., by polarizing the nuclear spins \cite{Eble_PRB06,Hogele_PRL12,Gullans_PRL10} in one QD to obtain an effective local magnetic field through the Overhauser term of the hyperfine interaction, $I_z s_z$ or via the use of micromagnets \cite{Shin_PRL10,Sigillito_arxiv19}, which would provide a more deterministic approach.

\begin{figure}[htp]
	\centering
\includegraphics[width=0.5\columnwidth]{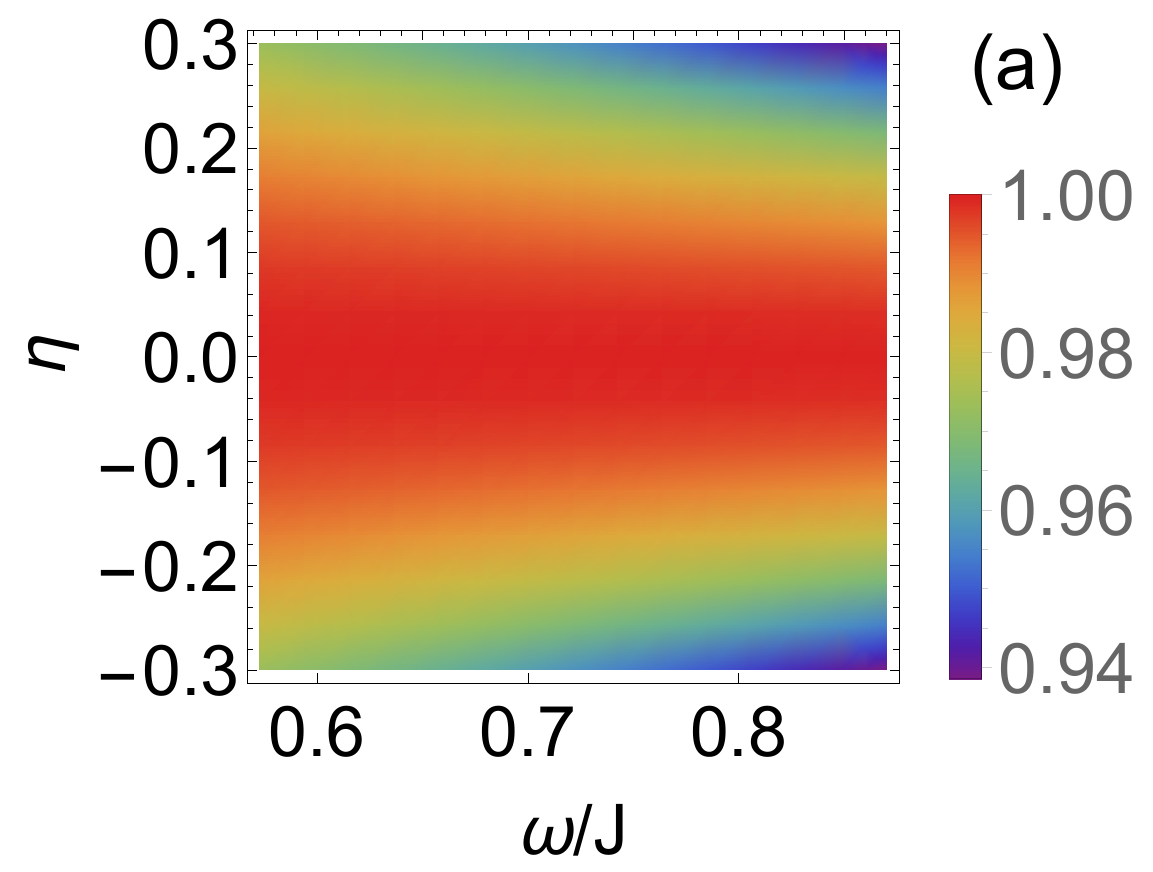}
\includegraphics[width=0.48\columnwidth]{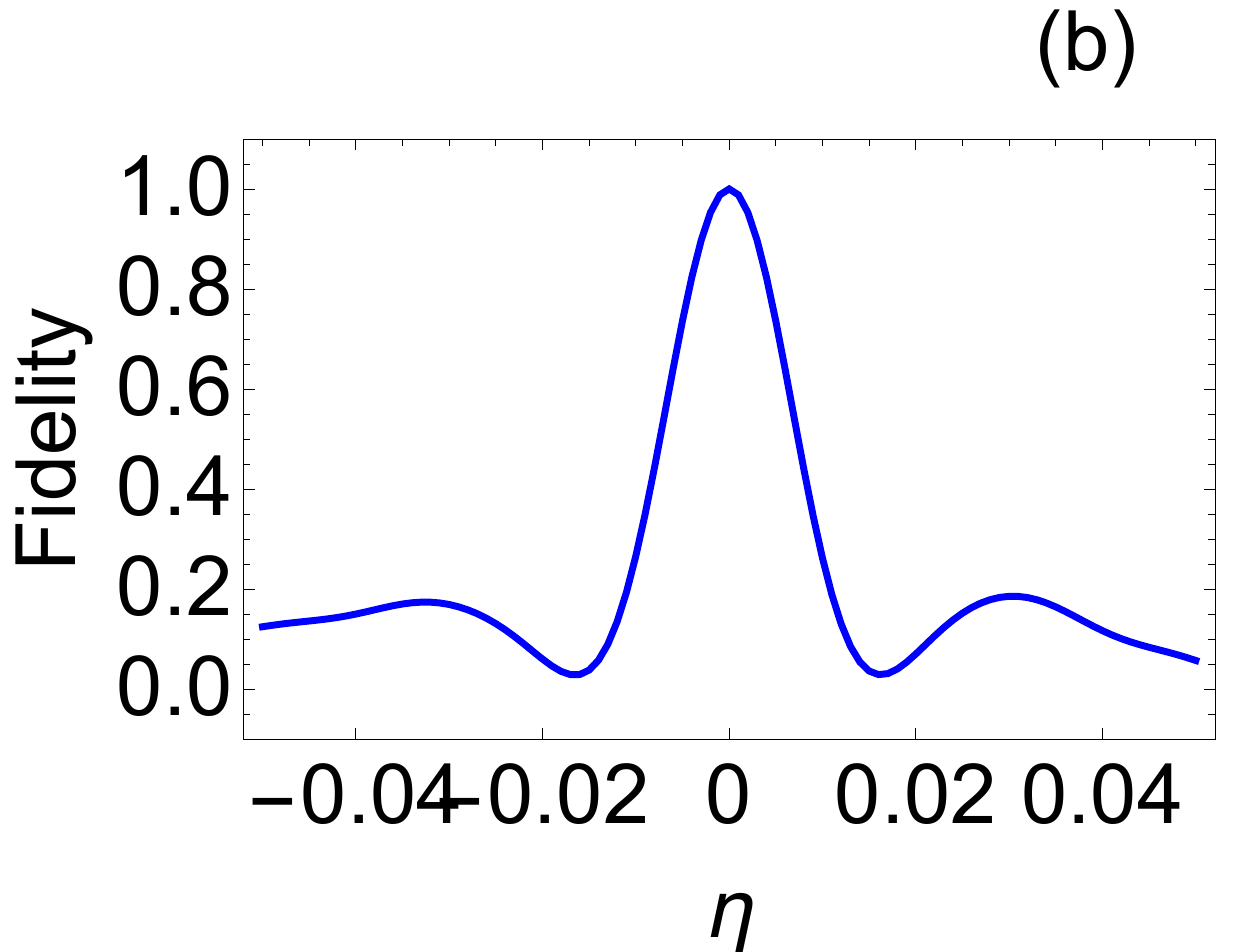}
\caption{Fidelity of pulse sequence (a) that provides the $x$ rotations in the equal Zeeman case as a function of the Zeeman frequency in units of $J$ and the percentage error $\eta$ in Zeeman frequency and (b) for the unequal Zeeman case as a function of the error $\eta$ in the ratio of the Zeeman frequencies. }
	\label{errorplot}
\end{figure}

\emph{Error analysis}--We now address how well the sequences presented above perform in the presence of errors, such as uncertainty in the system parameters. In the equal Zeeman case, we consider an error, $\omega {\rightarrow}\omega (1{+}\eta)$ and plot the fidelity of the pulse sequence of Eq.~(\ref{xrotfromz}), defined as $\mathcal{F}{=}|Tr(U U^\dag_{ideal})/4|^2$, as a function of the ratio $\omega/J$ for the allowed regime of interest and as a function of $\eta$, Fig.~\ref{errorplot}(a). We find that the fidelity is robust, even for high percentage of error in $\omega$. The remaining operations do not depend on $\omega$, so its fluctuations will not affect other parts of the full sequence.

In the unequal Zeeman case, we consider an error in the Zeeman ratio discussed above, $\omega_1/\omega_2{=}1.18+\eta$. The fidelity is shown in Fig.~\ref{errorplot}(b) and it depends more sensitively on the error. Such an error in $\eta$ could occur from fluctuations in the Overhauser field, which has typical widths on the order of 10s of MHz \cite{Bluhm_PRL10,Ethier-Majcher_PRL17}. Narrowing the nuclear spin distribution \cite{Bluhm_PRL10,Ethier-Majcher_PRL17} to ~1MHz (corresponding to $\eta\sim$ 0.001) would guarantee high fidelity. Despite the higher sensitivity to error, this regime has the advantage of not requiring a rotation about $x$ at all.
%\begin{figure}[htp]
%	\centering
%	\includegraphics[width=5cm]{errorplotunequal.pdf}
%	\caption{Fidelity of pulse sequence for the unequal Zeeman case as a function of the error $\eta$ in the ratio of the Zeeman frequencies.}
%	\label{errorplotuneq}
%\end{figure}

The $z$ gates are assumed to be instantaneous compared to the other timescales in the system. This is an excellent approximation, as faster pulses lead to a polarization selection rule for circularly polarized light, which in turn leads to higher fidelities \cite{Economou_PRB06}.

Additional sources of error are the finite trion lifetime and the finite spin coherence time. The implicit assumption has been that the QDs emit photons immediately after excitation, which requires that the spontaneous emission time be much faster than the Larmor precession periods and the timescale of the exchange interaction. In addition, the total pulse sequence should be much shorter than the spin coherence time so that a large enough cluster can be generated before the spin decoheres. The spontaneous emission time in free space is $\sim$1 ns, and it can be made faster through the Purcell effect by embedding the QD into a cavity \cite{Birowosuto_scirep12,Babinec_arxiv14, Vora_NC15}. Then an emission timescale on the order of 100 ps can be achieved \cite{Birowosuto_scirep12,Babinec_arxiv14}. Importantly, coupling to a cavity still allows for optical spin rotations by using off-resonant pulses \cite{Carter_NPhot13}. Ref. \cite{Lindner_PRL09} showed that the ratio of the the Zeeman frequency over the spontaneous emission rate can be as high as 10-20\% with reasonably low errors. This constrains the Larmor period and exchange interaction timescale to be on the order of 10 ns (1 ns) or longer for free space (cavity-mediated) emission. The sequences of free evolution and pulses have a duration roughly given by several times $\pi/J$. Taking $J{\sim}\omega_j{\sim} 2\pi\times 1$ GHz, each period should be ${\sim}$20 ns. The coherence time of the electron spin $T_2$ is several $\mu$s in free-induction decay and can be extended using decoupling sequences. Based on these values we estimate we can obtain a cluster state of size at least 2$\times$100, an order of magnitude larger than the state of the art.

In conclusion, we developed a method to generate a large 2D entangled photonic cluster state using coupled emitters. We showed in detail how this would work in a QD molecule with current experimental capabilities. Our approach can be adapted to other systems, including point defects, trapped ions, etc. Adding emitters to the system would increase the cluster state beyond two photons in the vertical direction. This could be done with a chain of emitters and decoupling to select at any one time coupling between only two neighboring emitters.

\emph{Acknowledgements}--We thank E. Barnes and S. Carter for useful discussions. SEE acknowledges support from NSF (Grant
No. 1741656).

%\bibliography{bibliography_CS_always_on}

%

\end{document}